\documentclass[12pt]{JHEP3}
\usepackage{amsmath,amsfonts,amssymb}
\usepackage{verbatim,graphicx,float}

\title{Observational hints on the Big Bounce}

\author{Jakub Mielczarek$^a$\footnote{jakub.mielczarek@uj.edu.pl}, 
Micha{\l} Kamionka$^{b}$\footnote{kamionka@astro.uni.wroc.pl}, 
Aleksandra Kurek$^{a}$\footnote{kurek@oa.uj.edu.pl} 
\ \ \ \ \ \ \ \ \ \ \ \ \ \ \ \ \ \ \ \
and Marek Szyd{\l}owski$^{ac}$\footnote{uoszydlo@cyf-kr.edu.pl}\\
$^a$ Astronomical Observatory, Jagiellonian University,\\ 
30-244 Krak\'ow, Orla 171, Poland\\
$^b$ Astronomical Institute, Wroc{\l}aw University\\
51-622 Wroc{\l}aw, Kopernika 11, Poland\\
$^c$ Mark Kac Complex Systems Research Centre, Jagiellonian University,\\
Reymonta 4, 30-059 Krak{\'o}w, Poland
}

\abstract{In this paper we study possible observational consequences of the 
bouncing cosmology. We consider a model where a phase of inflation is preceded 
by a cosmic bounce. While we consider in this paper only that the bounce is 
due to loop quantum gravity, most of the results presented here can be applied 
for different bouncing cosmologies. We concentrate on the scenario where 
the scalar field, as the result of contraction of the universe, is driven from the 
bottom of the potential well. The field is amplified, and finally the phase of the 
standard slow-roll inflation is realized. Such an evolution modifies the standard 
inflationary spectrum of perturbations by the additional oscillations and damping
on the large scales. We extract the parameters of the model from the observations
of the cosmic microwave background radiation. In particular, the value of 
inflaton mass is equal to  $m=(2.6 \pm 0.6) \cdot 10^{13}$ GeV. In our 
considerations we base on the seven years of observations made by the 
WMAP satellite. We propose the new observational consistency check for
the phase of slow-roll inflation. We investigate the conditions which have to 
be fulfilled to make the observations of the Big Bounce effects possible. 
We translate them to the requirements on the parameters of the model and
then put the observational constraints on the model. Based on assumption 
usually made in loop quantum cosmology, the Barbero-Immirzi parameter 
was shown to be constrained by $\gamma<1100$ from the cosmological 
observations. We have compared the Big Bounce model with the 
standard Big Bang scenario and showed  that the present observational data 
is not informative enough to distinguish these models.
}

\begin{document}

\section{Introduction} \label{Intro}

The observations of the cosmic microwave background (CMB) radiation indicate
that the power spectrum of primordial scalar perturbations is in the broad range
\emph{nearly} scale-invariant. Therefore, the spectrum can be written in the 
power-law form
\begin{equation}
\mathcal{P}_{\text{s}}(k) = A_{\text{s}} \left( \frac{k}{k_0}\right)^{n_{\text{s}}-1}, 
\label{powerspectrum}
\end{equation}
where the spectral index $n_{\text{s}}$ is close to unity. Here, $A_{\text{s}}$ is 
an amplitude of the scalar perturbations and $k_0$ is the so-called pivot number. 
The case $n_{\text{s}}=1$ corresponds to the scale-invariant Harrison-Zeldovich 
spectrum. The observed spectrum is almost of this type, namely it is little red-shifted
($n_{\text{s}} \lesssim 1$). In particular, the seven years of observations made by the 
WMAP satellite \cite{Komatsu:2010fb} indicate that $n_{\text{s}} = 0.963\pm 0.012$ (68\% CL).

The spectrum in the form discussed above can be explained by the phase of cosmic
inflation (see e.g. \cite{Linde:2007fr}). This phase can be driven by the self-interacting scalar field, the so-called 
inflaton field. In the most conservative approach the inflation can be driven by a single
massive scalar field. This case will be considered in this paper. In this model, a nearly
scale-invariant spectrum spectrum is generated during the slow-roll phase. In a more
general case the inflation can be driven with the different potentials. However, the 
potentials other than massive, lead to the non-Gaussian structure of the cosmic 
primordial perturbations what can be constrained by the CMB observations \cite{Bartolo:2004if}. 
Since the CMB anisotropies do not indicate any non-Gaussian signatures, the massive 
potential is somehow privileged. However, with the present sensitivity on these kind
of effects, some of the other potentials are still allowed.  Other models, as multi-field
inflation are also possible to be realized. In this paper, we consider the simplest possible 
realization of the inflation which is given by the single massive scalar field.

The weak point of the slow-roll inflationary scenario is that it requires some special
initial conditions. Namely, the field has to start its evolution not from the bottom of 
potential well but from the position which is far from its center. In the classical model
there is no mechanism to drive this field up the potential well. However, it has been 
recently pointed out \cite{Mielczarek:2009zw} that the phase of a quantum bounce 
can drive the inflaton field up the potential well and set the proper initial conditions  
for the slow-roll phase. The studies were performed within the loop quantum cosmology
(LQC) \cite{Bojowald:2008zzb,Ashtekar:2008zu}, however the mechanism is generic 
for all models with the bouncing phase. In the framework of LQC this issue has been 
studied recently in \cite{Mielczarek:2009zw,Ashtekar:2009mm,Chiou:2010nd,Mielczarek:2010bh}.

In the framework of LQC the classical dynamics of the universe is significantly 
modified when the energy density approaches the Planck energy density. These 
effects of the quantum gravitational modification can be introduced as corrections 
to the classical equations of motion. In particular, the modified Friedmann equation
takes the form
\begin{equation}
\left(\frac{1}{a}\frac{da}{dt}\right)^2=
\frac{8\pi}{3m^2_{\text{Pl}}} \rho\left(1-\frac{\rho}{\rho_{\text{c}}} \right), 
\label{Friedmann}
\end{equation}
where the critical energy density is defined as follows
\begin{equation}
\rho_{\text{c}}=\frac{\sqrt{3}}{16\pi^2\gamma^3}\rho_{\text{Pl}}, \label{rhoc}
\end{equation}
where $\rho_{\text{Pl}}:=m^4_{\text{Pl}}$ and $m_{\text{Pl}}\approx1.22\cdot 10^{19}$ 
GeV is the Planck mass.  The above expression (\ref{rhoc}) is crucial, because it relates 
the Barbero-Immirzi parameter $\gamma$ with the parameter $\rho_{\text{c}}$, which 
can be constrained observationally. This is a good example of expressing the 
phenomenological parameter, as $\rho_{\text{c}}$, in terms of parameters of the 
underlying theory. In this case the fundamental theory is loop quantum gravity (LQG)
with a free parameter $\gamma$. 

The value of Barbero-Immirzi  parameter is usually fixed from the considerations of 
the black hole entropy \cite{Ashtekar:1997yu}. In particular, the value $\gamma=0.239$ 
\cite{Meissner:2004ju} was derived what leads to $\rho_{\text{c}}=0.82 \ m^4_{\text{Pl}}$. 
The expression (\ref{rhoc}) is however not free from ambiguities. In particular, it bases on 
assumption that the area of the loop in LQC is equal to a gap of the area operator 
within LQG. This is not necessary true and therefore findings based on expression
(\ref{rhoc}) must be carried with a due care. In general, $\rho_{\text{c}}$ can be treated
as a free phenomenological parameter (see e.g. \cite{Dzierzak:2008dy,Malkiewicz:2009zd,
Mielczarek:2010rq}).  

\section{Inflation in LQC}

The global dynamics of the considered model was studied in \cite{Singh:2006im}.
It was shown there that the model possesses generic inflationary attractors.
However, the particular evolutionary paths can differ. Here we restrict our 
considerations to the one particular scenario where the energy density 
is dominated by the kinetic term at the bounce. This condition guarantee that the  
quantum back-reaction effects can be neglected and equation  (\ref{Friedmann}) can 
be applied. This issue will be discussed in more details at the end of this section.

In the considered evolutionary scenario, we initiate evolution from the contracting phase. 
The scalar field is initially placed at the bottom of the potential well, what seems to be 
the realistic and conservative assumption. The initial energy density is therefore contained 
in the kinetic part only. Because we begin the evolution at the low energy scales, the 
corresponding time derivative of the field is also small.  Another possible choice of initial 
conditions is given e.g. by saturating the Heisenberg uncertainty relation, then 
$\pi_{\phi}(t_0) \phi(t_0) ={\hslash}/2$ at some initial time $t_0$. 

The field begins its evolution from some tiny quantum fluctuations and starts to 
oscillate at the bottom of the potential with the time scale of oscillations proportional 
to $m^{-1}$. The dynamics of the field is governed by the unmodified, standard equation  
\begin{equation}
\ddot{\phi}+3H\dot{\phi}+m^2\phi=0 \label{KG}. 
\end{equation}
During the contraction, the Hubble factor is negative, and therefore the second term 
in equation (\ref{KG}) acts as \emph{anti-friction}. Moreover, while the universe contracts
the absolute value of $H$ increases. Therefore as an effect of growing \emph{anti-friction},
the oscillations are amplified. While approaching the bounce, the field becomes rapidly 
displaced from the equilibrium state. After that, the standard slow-roll inflation starts. 
We present the described evolution of the scalar field in Fig. \ref{Evolution1}.
\FIGURE{
\centering
\includegraphics[width=8cm,angle=0]{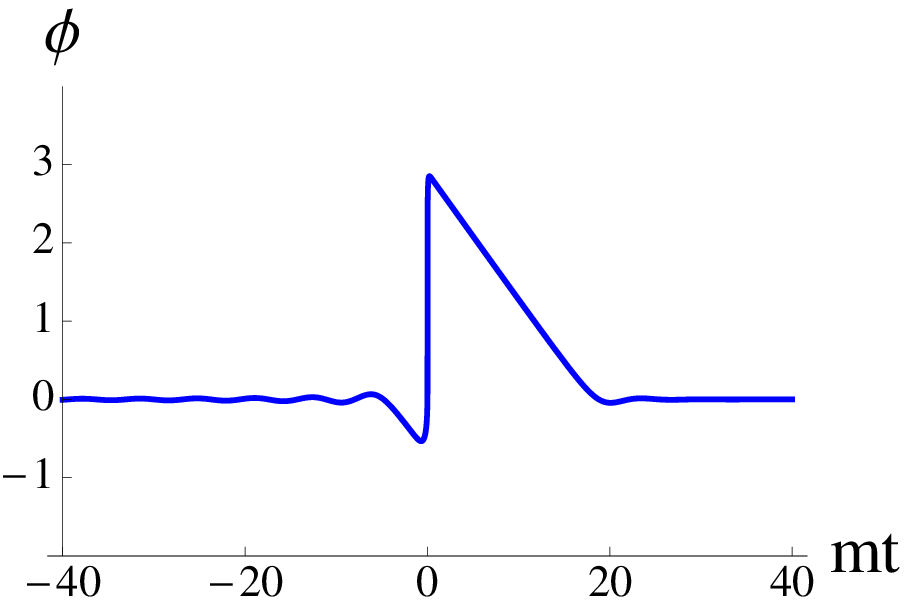}    
\caption{The \emph{shark fin} type evolution of the scalar field for $m=10^{-4}m_{\text{Pl}}$.
Here $\phi_{\text{max}} \approx 3 m_{\text{Pl}}$, what leads to the total number of 
$e$-foldings $N\approx 56$.}
\label{Evolution1}
}
The described evolution corresponds to the \emph{shark fin} scenario discussed in 
\cite{Mielczarek:2010bh}. During the evolution, the scale factor decreases in the 
pre-bounce stage. Since in this stage the field behaves effectively as a dust matter, 
the scale factor $a \propto |t|^{1/3}$.   The scale factor reaches its minimal value and then 
expands almost exponentially, during the slow-roll inflation. While the inflation ends,
the scale factor increases as $a \propto |t|^{1/3}$. The numerically computed evolution 
of the scale factor has been shown in Fig. \ref{Evolution2}.
\FIGURE{
\centering
\includegraphics[width=8cm,angle=0]{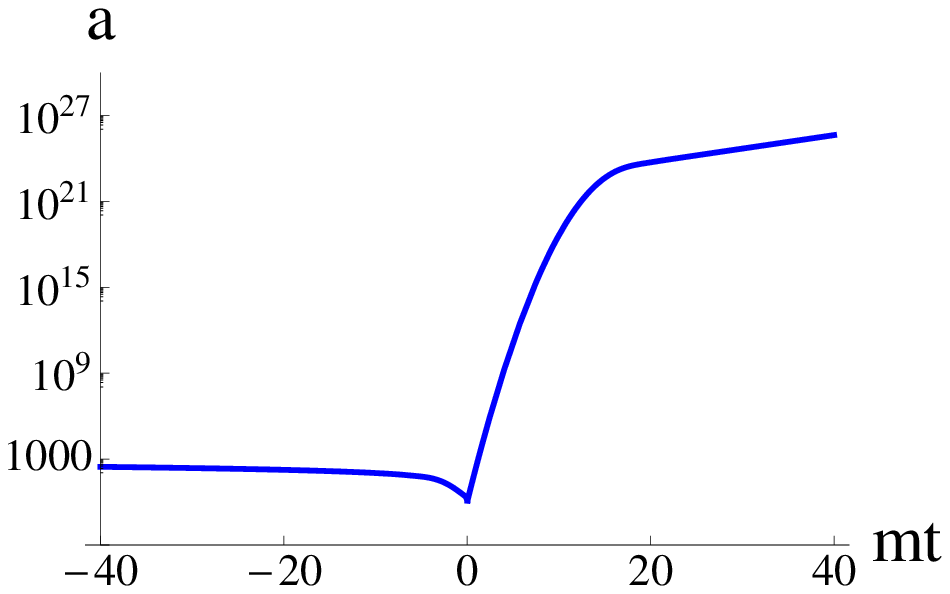}   
\caption{Evolution of the scale factor for $m=10^{-4}m_{\text{Pl}}$.}
\label{Evolution2}
}
We finish this section with some remarks on the quantum back-reaction effects. 
These effects become significant when the evolution of the higher moments 
influences behavior of the mean values of the quantum operators. In the free field 
case, the evolution of the mean values decouples from the evolution of the higher 
moments (for detailed discussion of this issue we refer to  \cite{Bojowald:2008pu}).
The back-reaction effects appear however in the presence of the potential of the scalar 
field and can significantly modify the effective dynamic. In particular, equation 
(\ref{Friedmann}) holds only when the quantum back-reaction effects can be neglected.
Otherwise this equation can be generalized to \cite{Bojowald:2008ec}
\begin{equation}
\left(\frac{1}{a}\frac{da}{dt}\right)^2=\frac{8\pi}{3m^2_{\text{Pl}}}\left[\rho\left(1-\frac{\rho_Q}{\rho_{\text{c}}} \right)
\pm\frac{1}{2}\sqrt{1-\frac{\rho_{Q}}{\rho_{\text{c}}}} \eta (\rho-P)+\frac{(\rho-P)^2}{2(\rho+P)}\eta^2 \right], \label{QBR}
\end{equation}
and the dynamics can be much more complicated that this discussed previously.
Here $\eta$ parameterizes strength of the quantum back-reaction effects. In the 
considered massive field case the energy density and pressure are respectively  
\begin{equation}
\rho= \frac{\dot{\phi}^2}{2}+\frac{m^2\phi^2}{2} \ \ \text{and} \ \ P= \frac{\dot{\phi}^2}{2}-\frac{m^2\phi^2}{2}.
\end{equation}
When $P=\rho$, the quantum back-reaction effects disappear,  equation (\ref{QBR}) simplify 
to  (\ref{Friedmann}). This corresponds to the free field case.  Therefore, while the energy density 
is dominated by the kinetic part, the quantum back-reaction effects can be neglected.
In Fig. \ref{density} we show an exemplary evolution of the energy density in the considered model.
We also present contributions from the kinetic and potential parts. 
\FIGURE{
\centering
\includegraphics[width=8cm,angle=0]{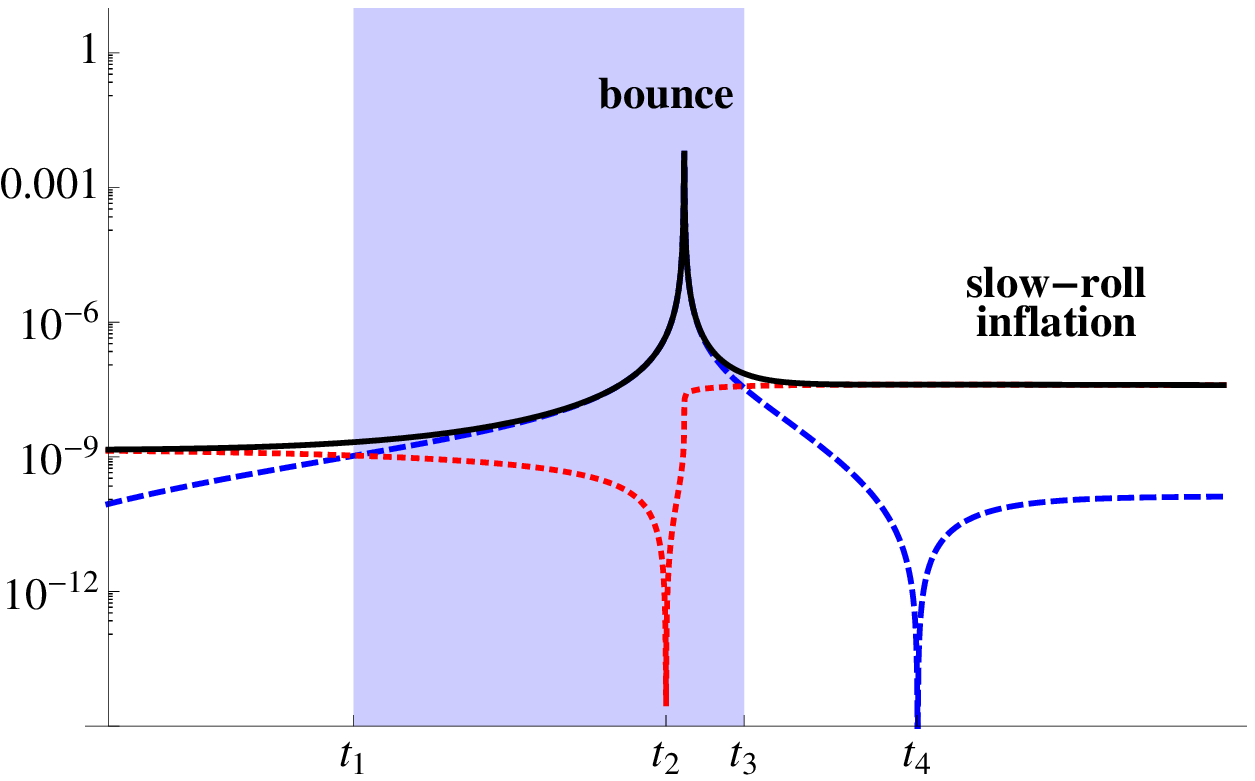}
\caption{The solid (black) curve represents the evolution of the total energy
density of the scalar field. The dashed (blue) curve represents the contribution
from the kinetic part. The dotted (red) curve represents the contribution from
the potential part. In the filled region the energy density is dominated by the
kinetic part. Here we have assumed $m=10^{-4}m_{\text{Pl}}$.}
\label{density}
}
The kinetic term $\frac{\dot{\phi}^2}{2}$ dominates the potential part $\frac{m^2\phi^2}{2}$ 
in the broad region around the bounce. This is the shadowed region from $t_1$ to $t_3$ in 
Fig. \ref{density}. Therefore approximation based on (\ref{Friedmann}) 
holds. For the densities $\rho \ll \rho_{\text{c}}$ the kinetic part can be dominated by the potential
part.  However, at these densities the quantum effects become unimportant and dynamics can
be approximated by the classical equations. Therefore condition $\frac{\dot{\phi}^2}{2} \gg 
\frac{m^2\phi^2}{2}$ should be fulfilled only in the vicinity of the bounce. If it is not, the  quantum 
back reactions must be taken into account.  At time $t_2$, the bounce takes place and the energy
density reaches its maximal value $\rho_{\text{c}}$. Later, the energy density decreases and holds 
at approximately constant value. This is the sign that the phase of inflation starts. Precisely, the 
inflation starts at time $t_4$  when the field turns round. At this point, the kinetic term falls to zero 
since the field stops for a moment. Thereafter, the field makes a slow-roll from the top of the potential well.
This part of evolution is almost purely classical and the quantum corrections can be neglected. 

\section{Observational hints on the slow-roll inflation}

Before we proceed to investigate the possible effects due to the bounce, we will firstly 
discuss the present observational hints regarding the inflation. It is crucial since probing 
the inflationary phase is more observationally available and give the chance to fix some 
parameters of the model. We will discuss here what we can already say about the 
slow-roll inflation model in light of the latest WMAP observations. We also propose the 
consistency check on the slow-roll inflationary scenario. 

The seven years of observations made by the WMAP satellite give the following values 
of the amplitude and spectral index of the scalar perturbations \cite{Komatsu:2010fb}
\begin{eqnarray}
A_{\text{s}} &=& 2.441^{+0.088}_{-0.092} \cdot 10^{-9},  \label{AsWMAP}\\ 
n_{\text{s}} &=& 0.963 \pm 0.012, \label{nsWMAP}
\end{eqnarray}
at the pivot scale $k_0=0.002\ \text{Mpc}^{-1}$ (see equation (\ref{powerspectrum})). 
The prediction from the slow-roll is the spectrum of scalar primordial perturbations in the form
\begin{equation}
\mathcal{P}_{\text{s}}(k) = \underbrace{\frac{1}{\pi \epsilon} 
\left(\frac{H}{m_{\text{Pl}}} \right)^2}_{:=S} \left( \frac{k}{aH}\right)^{n_{\text{s}}-1}, \label{spowersl}
\end{equation}
as well as the spectrum of the tensor perturbations (gravitational waves) in the form   
\begin{equation}
\mathcal{P}_{\text{t}}(k) = \underbrace{\frac{16}{\pi} 
\left(\frac{H}{m_{\text{Pl}}} \right)^2}_{:=T} \left( \frac{k}{aH}\right)^{n_{\text{t}}}.
\end{equation}   
Expressions for the scalar and tensor spectral indices are respectively
\begin{equation}
n_{\text{s}}=1+2\eta-6\epsilon,  \label{ns}
\end{equation}
and 
\begin{equation}
n_{\text{t}}=-2\epsilon,  \label{nt}
\end{equation}
where $\epsilon,\eta \ll 1$ are called slow-roll parameters. 
The $\eta$ and $\epsilon$ are  defined in the following way
\begin{eqnarray}
\epsilon &\equiv& \frac{m^2_{\text{Pl}}}{16\pi} \left(\frac{V'}{V} \right)^2 =
\frac{m^2_{\text{Pl}}}{4\pi} \frac{1}{\phi^2}, \label{epsilon} \\
\eta &\equiv& \frac{m^2_{\text{Pl}}}{8\pi} \left(\frac{V^{''}}{V} \right) = 
\frac{m^2_{\text{Pl}}}{4\pi} \frac{1}{\phi^2}, \label{eta}
\end{eqnarray}
so for the massive inflation field, $\eta=\epsilon$. Based on this and equation
(\ref{ns}) we find $\epsilon=(1-n_{\text{s}})/4$. With use of (\ref{nsWMAP}), this
gives us $\epsilon=0.010\pm 0.003$. Therefore, the slow-roll condition $\epsilon \ll 1$
is indeed fulfilled. Moreover, based on  (\ref{nsWMAP}), the tensor spectral index
is predicted to be 
\begin{equation}
n_{\text{t}}=\frac{n_{\text{s}}-1}{2}=-0.019 \pm 0.006.
\end{equation}

Above we have related observations of the spectral index  with the expression
predicted from the slow-roll inflation. It was straightforward since both, predicted 
spectrum (\ref{spowersl}) and spectrum used in fitting (\ref{powerspectrum}) had the
same power-law form.  However relating the fitted parameter $A_{\text{s}}$ with $S$ 
requires additional discussion. At the pivot scale $\mathcal{P}_{\text{s}}(k=k_0)=A_{\text{s}}$.
Moreover, we know that the inflationary spectrum at a given mode $k$ is formed 
when this mode crosses the horizon, namely when $k\simeq aH$. Afterward, the 
spectrum holds the form fixed at the horizon. Therefore at the given scale $k$, the spectrum is 
$\mathcal{P}_{\text{s}}(k=aH)=S$. Since $S$ decreases with time, the observed 
spectrum has the falling tendency (governed by the power law dependence). Based on
this, one can relate $\mathcal{P}_{\text{s}}(k=k_0)=\mathcal{P}_{\text{s}}(k=aH)$, 
what gives $A_{\text{s}}=S$. Therefore, $A_{\text{s}}$ gives us the value of $S$ 
at the point when the mode, which is at present equal to $k_0$, had crossed the 
horizon during the inflation. This observation will be crucial for the later considerations. 
  
In order to quantify the contribution from the tensor modes it is convenient to consider
the ratio
\begin{equation}
r \equiv \frac{\mathcal{P}_{\text{t}}(k=k_0)}{\mathcal{P}_{\text{s}}(k=k_0)}
=\frac{T}{S}= 16\epsilon = 4(1-n_{\text{s}}) = 0.15 \pm 0.05, \label{r}
\end{equation}
where in the last equality we have used the WMAP results (\ref{nsWMAP}). This 
result is consistent with the present constraints on the contribution from the tensor
modes  
\begin{eqnarray}
r &<& 2.1 \ \text{at} \ 95\% \ \text{CL}  \ (\text{WMAP-7}, \cite{Komatsu:2010fb}), \\
r &<& 0.73 \ \text{at} \ 95\% \ \text{CL}  \ (\text{BICEP},\cite{Chiang:2009xsa}).
\end{eqnarray}
Moreover, the value predicted in (\ref{r}) is above the observational threshold on
detection of the PLANCK satellite \cite{:2006uk}. Therefore, if the predictions 
of the slow-roll inflation are correct, the tensor modes should be observed by 
the PLANCK mission.  
  
The next parameter that can be computed is the value of the scalar field. 
Combining  (\ref{epsilon}), (\ref{eta}) and (\ref{ns}) we get 
\begin{equation}
\phi_{\text{obs}} = \frac{m_{\text{Pl}}}{\sqrt{\pi(1-n_{\text{s}})}}=2.9 \pm 0.5 m_{\text{Pl}}.  
\label{phiobs}
\end{equation}
At this value of field, the observed structures were created \footnote{Precisely it is 
the value of $\phi$ at which mode which is at present equal to $k_0$, had crossed 
the horizon during the inflation.}. Therefore it can be treated as an lower limit on the 
maximal displacement of the scalar field. The maximal value of the scalar field is 
unbounded within the classical theory. However LQC puts the constrain on its value 
since the energy density is bounded by $\rho_{\text{c}}$. Based on this one find that  
\begin{equation}
|\phi |  \leq \frac{\sqrt{2\rho_{\text{c}}}}{m}. \label{philim}
\end{equation}

The found value of the inflaton field  (\ref{phiobs}) can be translated into the corresponding
$e$-folding number
\begin{equation}
N_{\text{obs}}\simeq 2\pi \frac{\phi^2_{\text{obs}}}{m^2_{\text{Pl}}}=\frac{2}{1-n_{\text{s}}}= 54 \pm 18.
\end{equation}
This is also not the total $e$-folding number for inflation, but only the lower limit
on its value. The total $e$-folding number is, in LQC, constrained by 
\begin{equation}
N \leq \frac{4\pi\rho_{\text{c}}}{m^2_{\text{Pl}}m^2},
\end{equation}
what bases on (\ref{philim}). Finally one can also derive the mass of inflaton field. Namely  
\begin{eqnarray}
m &\simeq& m_{\text{Pl}} \frac{1}{4}\sqrt{3\pi A_{\text{s}}} (1-n_{\text{s}}) \nonumber \\
&=& (1.4 \pm 0.5) \cdot 10^{-6} m_{\text{Pl}} \nonumber \\
&=& (2.6 \pm 0.6) \cdot 10^{13} \text{GeV}.
\end{eqnarray}
Therefore, one of the parameters of the model is fixed. The remaining  parameter 
$\rho_{\text{c}}$ is however harder to determinate. We will discuss the present 
observational constraint on $\rho_{\text{c}}$ in Sec. \ref{BBounce}. We stress that 
it was possible to determinate the value of the parameter $m$ basing only on the 
observational effects of inflation. It was not necessary to introduce any LQC effects here, 
because they were negligible during the phase of inflation. In other words, the slow-roll inflation is 
the classical (and observationally available) part of the considered \emph{shark fin} scenario.

The usual consistency check of the inflationary models bases on expressing  
of the tensor-to-scalar $r$ in terms of others (measured) parameters of the model. 
In case of the slow-roll inflation, we have considered it in equation (\ref{r}) and we have shown
that the derived value of $r$ places within the observational bound.  Here, we propose 
an additional consistency check for inflation. This new consistency relation requires 
however information about the duration of the reheating phase. Alternatively, the 
method can be used to put a constraint on the phase of reheating after inflation. 

The consistency check base on the fact that the modes created at the particular point of inflation 
where $\phi=\phi_{\text{obs}}$, correspond to the present pivot scale at which the amplitude of 
perturbations was computed. These particular modes have the size of horizon when created from the quantum fluctuations.  
Therefore, at this particular point 
\begin{equation}
\lambda_{\text{H}}=2\pi \frac{a}{k}\simeq\frac{1}{H}=
\frac{2}{\sqrt{\pi(1-n_{\text{s}})A_{\text{s}}}}=1.2 \cdot 10^{5} l_{\text{Pl}}.
\end{equation}
In turn, the present pivot scale is equal to 
\begin{equation}
\lambda_0 =\frac{2\pi}{k_0} = 3.14 \cdot 10^{3}\ \text{Mpc}.
\end{equation}
Based on this, one can find the total increase of the scale factor from the point 
at which $\phi=\phi_{\text{obs}}$, till now. We obtain the value
\begin{equation}
\Delta_{\text{tot}}:=\frac{a_0}{a_{\text{H}}}=\frac{\lambda_0}{\lambda_{\text{H}}} 
= 5\cdot 10^{55}. \label{Dtot}
\end{equation}

There is also another way to compute this quantity. Namely, starting 
from $\phi=\phi_{\text{obs}}$, the length $\lambda_{\text{H}}$ grows thereafter
till the end of inflation, across the reheating, radiation domination phase, matter
domination phase until now. It is hard to precisely determinate the increase of 
the scale factor at this whole evolution. In particular, because we do not know 
the duration of reheating phase and the duration till the end of inflation was 
determined with the significant uncertainty. Therefore we can perform only a 
raw approximation of the total increase of the scale factor. 

Let us collect the particular contributions starting from the present and going
backward:
\begin{itemize}
\item Matter era. The period from recombination till now. 
$\Delta_{\text{mat}}:=1+z_{\text{dec}}\simeq 10^3$

\item Radiation era. The period from the end of reheating till the 
recombination. $\Delta_{\text{rad}}:=\frac{T_{\text{GUT}} }{T_{\text{dec}}}=
\frac{10^{14}\ \text{GeV}}{0.2\ \text{eV}} 
\simeq 5\cdot 10^{23}$

\item Reheating (see e.g. \cite{Kofman:1994rk,Bassett:2005xm}). 
The period when the particles are created from the decaying 
inflaton field and the universe thermalizes. The corresponding quantity
$\Delta_{\text{reh}}$ is model dependent and should be fixed for the
particular scenario. For instance, for the instantaneous reheating 
$\Delta_{\text{reh}} \sim 1$. However, for the considered chaotic 
inflation the duration of reheating can be longer. We leave detailed 
considerations to this issue for the further studies, and now let the 
value of $\Delta_{\text{reh}}$ as a free parameter.

\item Inflation. The increase of the scale factor is equal to 
$\Delta_{\text{inf}}:= e^{N_{\text{obs}}} \simeq 3 \cdot 10^{23}$ 

\end{itemize} 

Based on this, the total increase of the scale factor is equal to 
\begin{equation}
\Delta_{\text{tot}} = \Delta_{\text{inf}}\Delta_{\text{reh}}\Delta_{\text{rad}}\Delta_{\text{mat}}. \label{Dt}
\end{equation}
The left side in equation (\ref{Dt}) is determined from  (\ref{Dtot}). Based on
the above relation one can e.g. try to determine duration of reheating. Namely, we have   
\begin{equation}
\Delta_{\text{reh}} = \frac{\Delta_{\text{tot}}}{ \Delta_{\text{inf}}\Delta_{\text{rad}}\Delta_{\text{mat}}} =
\frac{5\cdot 10^{55}}{1.5 \cdot 10^{51}} \simeq 3 \cdot 10^{4}.  
\end{equation}

On the other hand, in order to use (\ref{Dt}) to verify the model of inflation, 
the duration of reheating must be known from the theory. Then one can 
define the quantity
\begin{equation}
\theta:=\frac{\Delta_{\text{tot}}}{ \Delta_{\text{inf}}\Delta_{\text{reh}}
\Delta_{\text{rad}}\Delta_{\text{mat}}}  = 
\frac{\pi^{3/2}\sqrt{(1-n_{\text{s}})A_{\text{s}}}}{k_0 \exp \left(\frac{2}{1-n_\text{s}}\right) 
\Delta_{\text{reh}}\frac{T_{\text{GUT}} }{T_{\text{dec}}}(1+z_{\text{dec}})} \label{theta}
\end{equation}
This is consistency relation for the cosmological model with the slow-roll inflation. 
The meaningful cosmological should fulfill the condition $\theta \approx 1$. At
present, the application of (\ref{theta}) is limited due to the unknown factor 
$\Delta_{\text{reh}}$. However, it could be possible to determine this value 
basing on the found value of $m$ and the decay rate of the inflaton field. 
This issue requires however detailed studies, therefore we leave it to 
investigate elsewhere.

\section{Modified inflationary spectrum and the CMB} \label{CMB}

As it was shown in Introduction, the spectrum from the slow-roll inflation 
can be parametrized in the power-law form. However, the prior phase of 
a bounce should results with modification of this spectrum. The modifications
of the primordial scalar spectrum, were investigated in the numerous papers 
\cite{Cai:2008ed,Wands:2008tv,Novello:2008ra}. 
However, the studies were performed only when the evolution of the scalar 
modes hold the classical form. Within loop quantum cosmology, not only 
the dynamics of the background is modified but also the perturbations 
\cite{Bojowald:2006tm,Bojowald:2008jv}.      
In case of the tensor modes (gravitational waves), the form of these modifications
was studied in details \cite{Bojowald:2007cd,Mielczarek:2009vi,Grain:2009eg,Copeland:2008kz}. 
Based on this, the spectrum from the  \emph{shark fin} scenario 
considered has been recently found in \cite{Mielczarek:2010bh}.
In this paper both quantum corrections to the background as well to the perturbation
part were taken into account.  The case of the scalar modes is however more
problematic. It is because of the issue of quantum anomalies of the algebra 
of constrains. In case of the so-called inverse-volume corrections, this problem 
has been resolved \cite{Bojowald:2008gz}. However, in case of the holonomy corrections the 
anomaly free equations are still not available. Some preliminary attempts 
to investigate LQC effects on the scalar power spectrum were performed
in Ref. \cite{Artymowski:2008sc}. However, the evolution of modes was treated 
classically and the quantum effects were introduced by the influence on  
the matter part.  Also so attempts to derive holonomy corrected  equations 
on scalar modes were performed in Ref. \cite{Wu:2010wj}. However, the authors 
neglected the issue of the anomaly free algebra of constraint. Therefore 
derived equations can drive the system out of the surface of constraint and 
lead to  erroneous predictions. Therefore, the systematic analysis of the 
anomaly freedom in case of the holonomy corrected scalar perturbations 
remains to be done. 

Before the details of modifications of the scalar power spectra will be available, 
it is advisable to perform the phenomenological analysis of the possible impact 
of these effects on the CMB spectrum. This issue of impact of the LQC effects 
on the CMB spectrum was also discussed in Ref. \cite{Mielczarek:2009zw,Grain:2009kw,Barrau:2009fz}.
However, here we perform quantitative analysis in contrast of the much more 
qualitative discussion in the cited papers. We assume that the scalar power spectrum 
takes a form
\begin{eqnarray}
\mathcal{P}_{\text{s}} (k) &=& \Delta(k,k_*)A_{\text{s}} \left( \frac{k}{k_0}\right)^{n_{\text{s}}-1}. 
\label{spectrum}
\end{eqnarray}  
This is, in fact, the standard inflationary spectrum modified by the 
additional prefactor $\Delta(k,k_*)$. The bounce-factor $\Delta(k,k_*)$ can be written as
\begin{equation}
\Delta(k,k_*)=1-\frac{\sin\left(\frac{3k}{2k_*}\right)}{\left(\frac{3k}{2k_*} \right)}, \label{FD}
\end{equation}
which is the  simplified form of the expression found in \cite{Mielczarek:2009zw}.
The $k_*$ is a parameter of the model and its interpretation will be discussed later. 
The factor $\Delta(k,k_*)$ reflects typical modifications which appear in the 
bouncing cosmology. In the UV limit, $\lim_{k\rightarrow \infty}\Delta(k,k_*)=1$, 
therefore the spectrum (\ref{powerspectrum}) is recovered. In turn, in the IR limit,  $\lim_{k\rightarrow 0}\Delta(k,k_*)=0$, 
and  the spectrum is suppressed.  This behavior of the power spectrum is typical 
for the bouncing cosmologies. The two effects of the bounce are transparent: suppression 
on the low $k$ and the additional oscillations. In Fig. \ref{Delta} we show function 
$\Delta$ defined by equation (\ref{FD}). Instead of using the wavenumber $k$ we have translated it to the corresponding 
length $\lambda=\frac{2\pi}{k}$, respectively  $\lambda_*=\frac{2\pi}{k_*}$.
\FIGURE{
\centering
\includegraphics[width=8cm,angle=0]{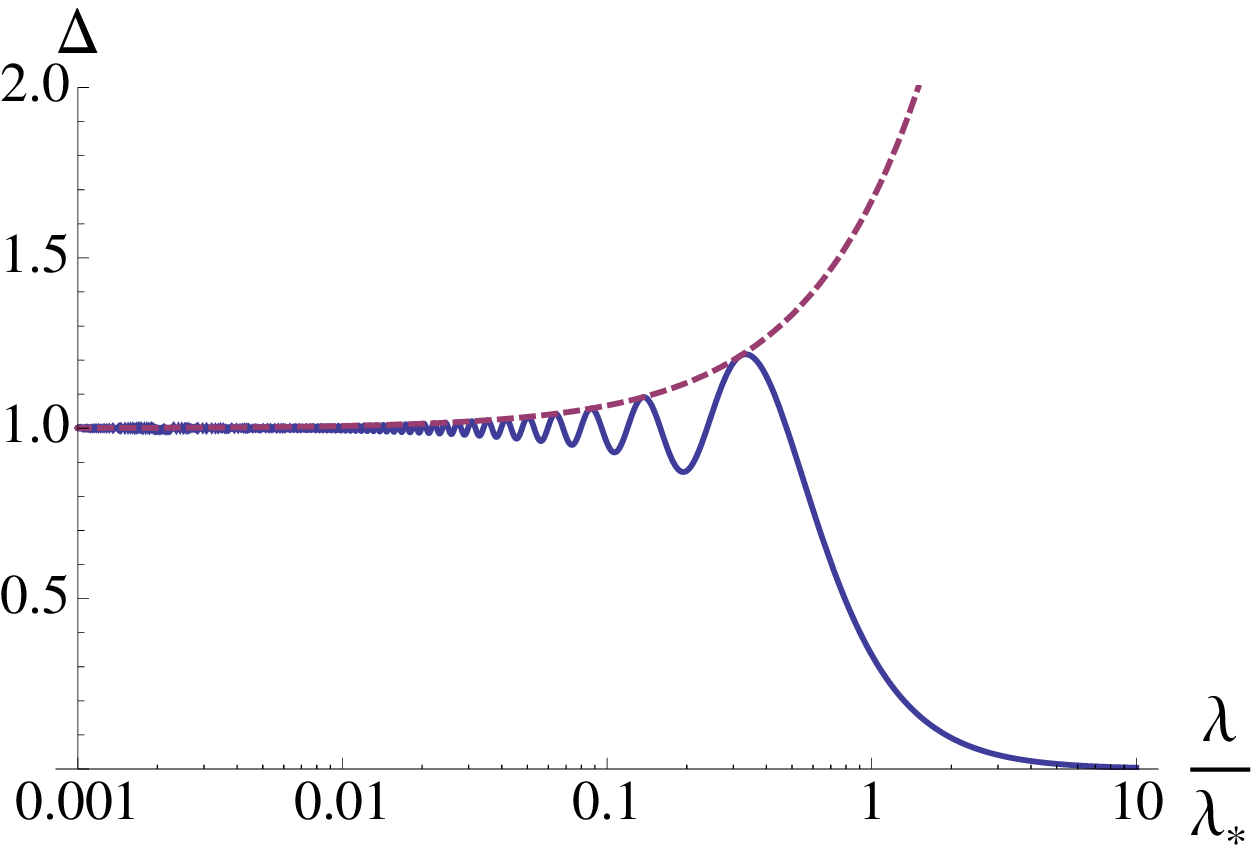}
\caption{Plot of function $\Delta$ defined by equation (\ref{FD}) (solid line). 
The dashed line represents approximation (\ref{appD}).}
\label{Delta}}
In Fig. \ref{Delta} we also show the function 
 \begin{equation}
\Delta(\lambda,\lambda_*) \approx 1+\frac{2}{3} \frac{\lambda}{\lambda_*}. \label{appD}
\end{equation}
This function measures the modification due to the oscillations for $\lambda/\lambda_*\ll 1$. 
At $\lambda/\lambda_*\approx1$ the spectrum becomes suppressed.
In the bouncing cosmology the length scale $\lambda_*$ can be related 
with the scale of horizon at the beginning of inflation. This issue was discussed in details 
in \cite{Mielczarek:2010bh}. Therefore if the present value of the scale factor is equal $a_0=1$ 
(as used in this paper), we have  $k_* \simeq a_i H_i$ where 
$a_i$ is the value of the scale factor at the beginning of inflation and 
$H_i$ is the value of the Hubble factor at the same time. Therefore if $k_*$ 
and $H_i$ could be measured, the total increase of the scale factor, from the
beginning of inflation till present, can be determined.  The value of $k_*$
and respectively $\lambda_*$ which is a scale of suppression in the spectrum 
is the crucial observational parameter of the bounce. In this paper we make 
an attempt of determining this value based on the observations of the CMB. 

As mentioned earlier, beside the effect of suppression, also oscillations of the 
spectrum are predicted within the bouncing cosmologies. This effects is much 
weaker that suppression, however is present also on the much smaller scales. 
This is important from the observational point of view. Namely, the length scale 
$\lambda_* = \frac{2\pi}{k_*}$  can be much larger than the present size of horizon 
($k/k_* \ll 1$). Then, the effect of suppression would be inaccessible observationally. 
However, some oscillations are still present on the sub-horizontal scales. Of course
the amplitude of these oscillations decreases while $k/k_* \gg 1$. If the scale 
$\lambda_*$ is however not much higher than 
the size of horizon, the effect  of sub-horizontal oscillations could be quite significant. 
The oscillations in the primordial power spectrum translate into the additional oscillation
in the spectrum of the CMB anisotropies (see e.g. \cite{Falciano:2008gt}).
For the small multipoles, this subtle effect  can
be dominated by the contribution from the cosmic variance. However, for the larger 
multipoles this effect can dominate. At these scales, improvement of the instrumental
resolution are still possible,  what gives the chance to, at least, put a stronger constrain 
on these effects.                      

In this section we confront the spectrum (\ref{spectrum}) with observed 
anisotropies of the cosmic microwave background radiation. We use 
the seven years of observations made by the WMAP satellite \cite{Komatsu:2010fb}.   
In the numerical calculations we use the publicly available CAMB code 
\cite{Lewis:1999bs} and Markov Chain Monte Carlo (MCMC) package 
CosmoMC \cite{Lewis:2002ah}  together with the CosmoClust code 
\cite{Shaw:2007jj} for computing the Bayesian evidence. The codes were 
suitably modified to investigate the spectrum (\ref{spectrum}).  
In computations, we take the standard cosmological parameters as follows
\begin{equation}
(H_0,\Omega_bh^2,\Omega_ch^2,\tau) = (70,0.0226,0.112,0.09) 
\end{equation}
and the pivot scale $k_0=0.05\ \text{Mpc}^{-1}$.

In Fig. \ref{CMB} we show spectrum of the CMB temperature anisotropies
obtained based on power spectrum (\ref{spectrum}).
\FIGURE{
\centering
\includegraphics[width=8cm,angle=270]{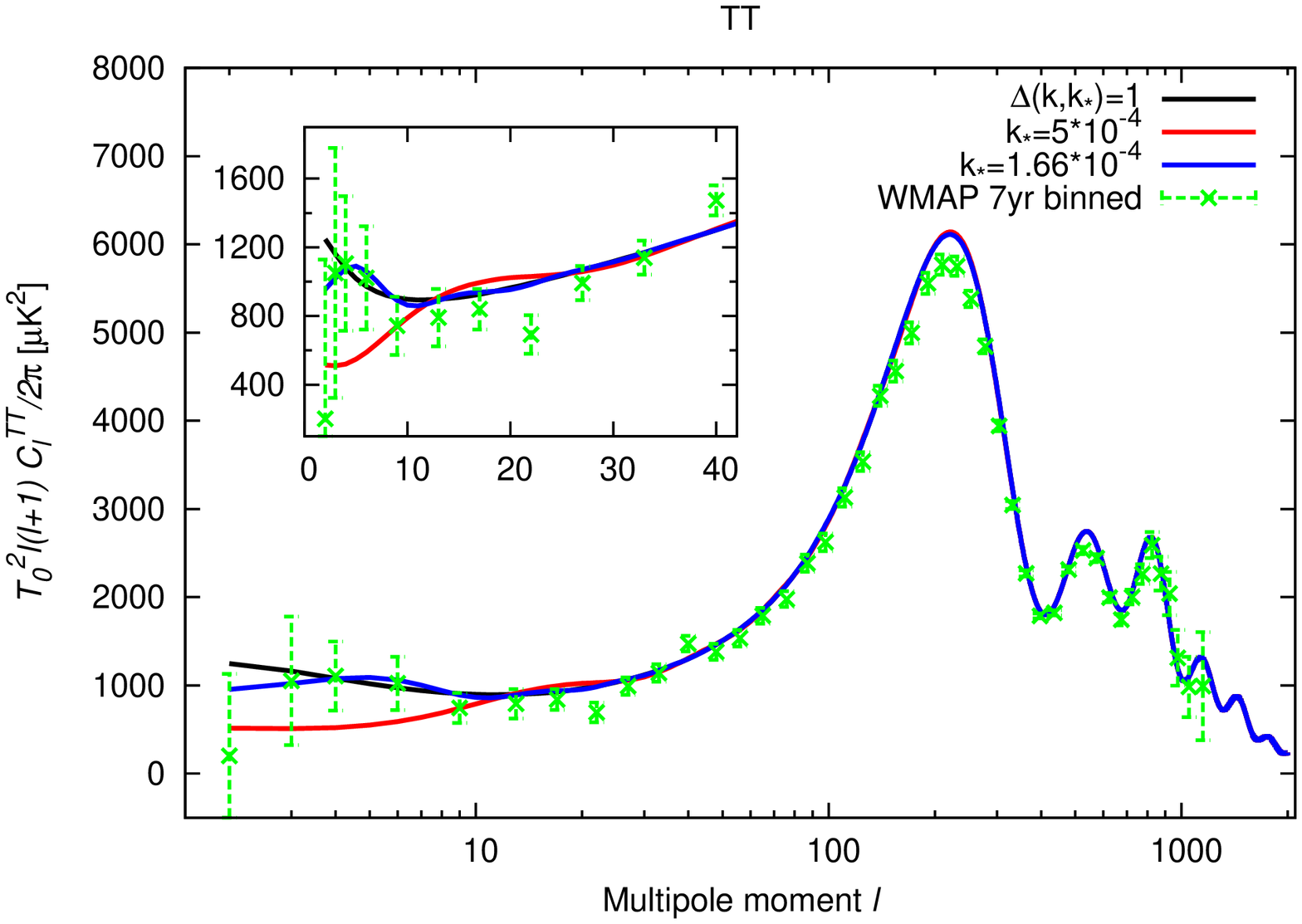}
\caption{Spectrum of the CMB anisotropy.}
\label{CMB}
}
The case $\Delta(k,k_*)=1 \ (k \rightarrow 0)$ corresponds to the classical case with no contribution
due to the bounce.  The blue line corresponds to the best fit case.  In this case, 
the modulations on  the low multipoles are well reproduced. This is due to the 
oscillations in the primordial power spectrum (\ref{spectrum}). This suggests 
that the effects of oscillations in the primordial power spectrum can be indeed 
studied basing on the CMB data. Perhaps the anomalous behavior of the CMB spectrum 
at $l \approx 20$ and $l \approx 40$ could be also explained by the oscillations within 
the bouncing scenario. However, not basing on the parametrization employed 
in this paper. The amplitude of oscillations on the lower scales must be higher 
than predicted by our model. 

We also find confidence intervals for the parameters of the model, namely on 
$A_{\text{s}}$, $n_{\text{s}}$ and $k_*$. In these computations we take into 
account the temperature anisotropy data (TT spectrum) as well as the 
polarization data (TE and EE spectra). We neglect a contribution from
the  tensor modes putting $\mathcal{P}_{\text{t}}=0$. We show the obtained 
confidence intervals in Fig. \ref{prob}.
\FIGURE{
\centering
\includegraphics[width=12cm,angle=0]{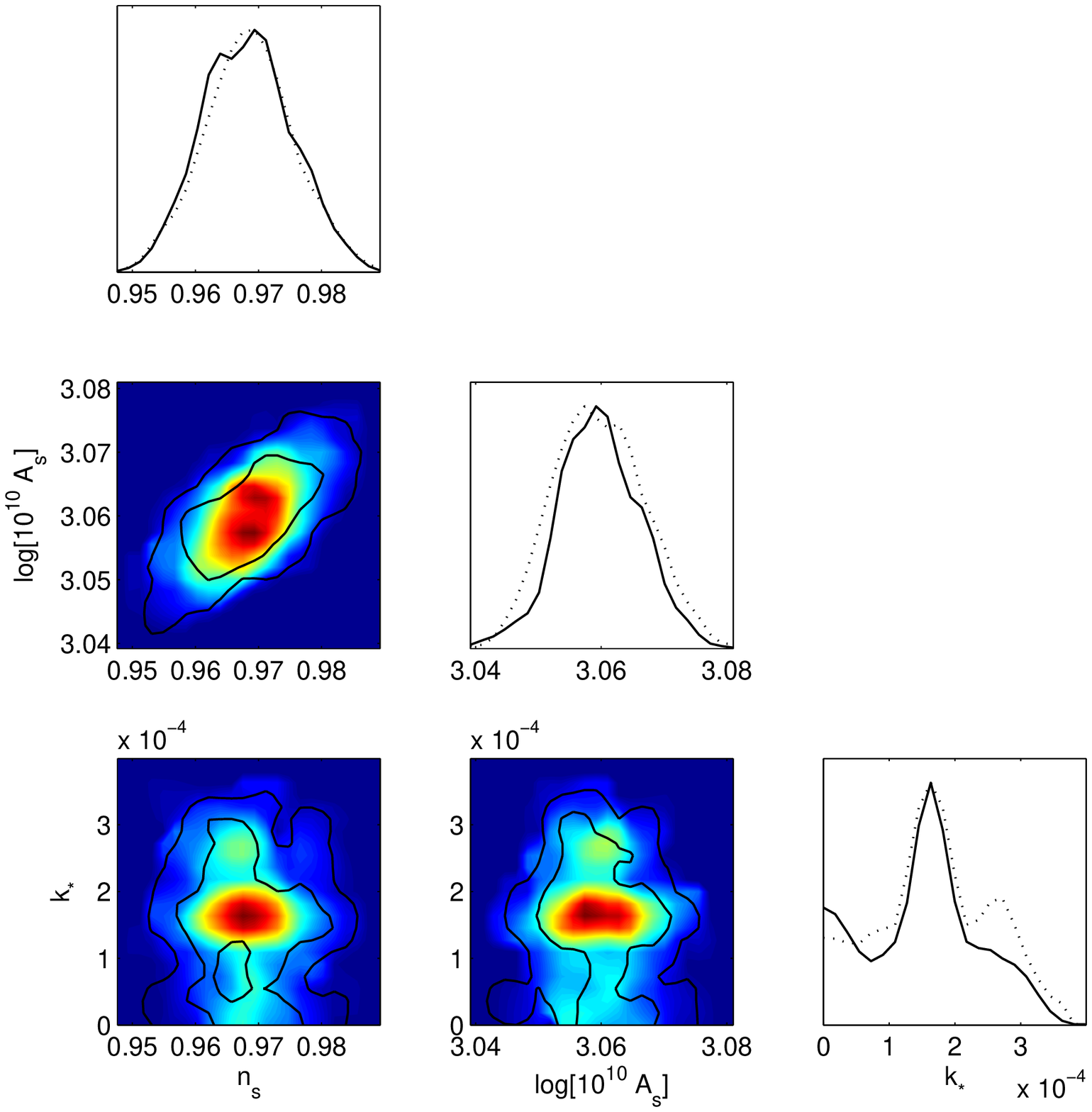}
\caption{Constraints for the parameters $A_{\text{s}}$, $n_{\text{s}}$ and $k_*$. 
2D plots: solid lines show the 68$\%$  and 95$\%$ confidence intervals. 
1D plots: dotted lines are mean likelihoods of samples, solid lines are 
marginalized probabilities.}
\label{prob}
}
As it can be seen from Fig. \ref{prob}, the parameters  $A_{\text{s}}$ are $n_{\text{s}}$
are constrained from the both sides. Based on the fit to the WMAP data we find 
\begin{eqnarray}
n_{\text{s}} &=& 0.97 \pm 0.07, \nonumber \\
A_{\text{s}} &=& 2.1 \cdot 10^{-9} \pm 0.1 \cdot 10^{-9}. \nonumber
\end{eqnarray}
These results are in agreement with (\ref{nsWMAP}) and  (\ref{AsWMAP}).
However it must be pointed out that they were computed at the different pivot 
scales.

The parameter $k_*$  has the upper constraint,  however it is unbounded from 
below (large scales). It could be expected, since there is 
no observational data on the largest (greater than the Hubble radius) scales to 
get the upper constrain the parameter. Nevertheless some particular value of 
$k_*$ is privileged what leads to the peak in the probability distribution. Based 
on the fit to the WMAP data we have obtained following values of this parameter:
\begin{equation}
k_* = 1.7 \cdot 10^{-4} \pm 0.8 \cdot 10^{-4} \  [\text{Mpc}^{-1}]. \nonumber
\end{equation}
The length scale corresponding to $k_*$ is equal to   
\begin{equation}
\lambda_*=\frac{2\pi}{k_*}  \approx 4 \cdot10^4\ \text{Mpc}.
\end{equation}

\section{Big Bang vs. Big Bounce}

In this section we compare the model with suppression with the standard
inflationary model. The suppression appears generically within the Big Bounce 
cosmology. In turn, the slow-roll inflation in the standard Big Bang scenario
does not lead to any suppression. While the suppression introduces a new 
length scale, the model with suppression has one more parameter in comparison
with the standard case.  The considered models are:
 \\

$H_1$ -- The slow-roll inflation within the Big Bang cosmology (spectrum given 
by Eq. \ref{powerspectrum}). This model has two parameters $A_{\text{s}}$ and $n_{\text{s}}$. 

$H_2$ -- The slow-roll inflation within the Big Bounce cosmology (spectrum 
given by Eq. \ref{spectrum}). This model has three parameters $A_{\text{s}}$, $n_{\text{s}}$ and $k_*$.
\\
 
In the Bayesian approach to model comparison the best model has the largest
value of the so-called posterior probability in the light of data, which is defined 
in the following way \cite{J61}:
\begin{equation}
P(H_i|D)=\frac{P(D|H_i) P(H_i)}{P(D)}.
\end{equation}
The $H_i$ stands for considered model and $D$ denotes data used in analysis. 
$P(H_i)$ is the prior probability for the model under investigation, which should 
reflect all information which we have about it before the analysis with the data 
$D$, that comes from theoretical investigations, or from analysis with other data
sets. In particular, if we have no foundation to favor of one model over another 
one, which is usually the case, we take equal values of $P(H_i)$ for all considered 
models. $P(D|H_i)$ is the marginalized likelihood function over the allowed 
parameters range, which we called evidence and is given by  
\begin{equation}
E_i\equiv P(D|H_i)=\int d\hat{\theta} L(\hat{\theta}) P(\hat{\theta}|H_i). 
\end{equation}
The $\hat{\theta}$ denotes vector of model parameters, $L(\hat{\theta})$ is the 
likelihood function for considered model and $P(\hat{\theta}|H_i)$ is the prior 
probability distribution function for model parameters. 

It is convenient to consider the ratio of models probabilities, which is reduced 
to the evidence ratio (so called Bayes factor) when all considered models have 
equal prior probabilities:
\begin{equation}
B_{ij}=\frac{E_i}{E_j}.
\end{equation}
Their values give us information about the strength of evidence in favor of better 
model \cite{RT08}: 
if $0< \ln B <1 $ we could not give conclusive answer, 
if $1< \ln B < 2.5$ there is weak evidence,
if $2.5 < \ln B < 5$ the evidence is moderate,
and for $\ln B >5$ evidence is strong.

The values of evidence for two alternative models of primordial
perturbation spectrum was calculated with the help of CosmoClust
code, which was introduced by \cite{Shaw:2007jj} as a part of CosmoMC code.
We have based on anisotropy (TT) and polarization (TE, EE) data 
from the WMAP satellite. In computations, we have neglected the contribution
from the tensor power spectrum (we set $\mathcal{P}_{\text{t}}(k)=0$).

We assume that models are equally probable ($P(H_1)=P(H_2)=1/2$). We 
consider flat prior probability distribution functions for unknown parameters 
in the following ranges: $A_{\text{s}} \in [1.5 \cdot10^{-9}, 5.5 \cdot 10^{-9}]$, 
$n_{\text{s}} \in [0.5, 1.5]$, $k_* \in [10^{-6},10^{-3}]$. The value of logarithm 
of the Bayes factor which was obtained in the analysis, i.e. 
\begin{equation}
\ln (E_1/E_2) = \ln B_{12} =0.2 \pm 0.6, 
\end{equation}
does not give a conclusive answer. The data was
not informative enough to distinguish these models.  Therefore, in the light of 
the recent WMAP data the Big Bang and Big Bounce cosmologies are 
indistinguishable. The Big Bounce predictions are not in conflict with the 
observational data.  Moreover, beside the fact that the Big Bounce model has
one more parameter $k_*$, the obtained evidence is comparable with the 
Big Bang case. 

The above result  was obtained with use of the CosmoClust code
which bases on the nested sampling method \cite{Mukherjee:2005wg}. 
This method was applied also in the CosmoNest code 
\cite{Parkinson:2006ku}. The computations with use of
 CosmoNest gives $\ln B_{12} = 1.1 \pm 0.2$. Therefore a
week evidence for Big Bang model is obtained. However, the  
CosmoNest was designed only for the case of the unimodal likelihood functions.
In turn, the CosmoClust code extends to the case of the multi-modal 
likelihood functions. As it is clear from the bottom right panel in Fig. \ref{prob},
the considered likelihood function (dotted line) is bimodal in the subspace 
$k_*$. The first peak is located at $k_* \sim 1.5 \cdot 10^{-4} \text{Mpc}^{-1}$ while 
the second at $k_* \sim 2.5 \cdot 10^{-4} \text{Mpc}^{-1}$. Therefore
the results from CosmoClust are more relevant for our model. The 
CosmoNest samples only around the highest peak, neglecting the 
contribution from the smaller one. Because of this, the observed 
discrepancy between  the CosmoClust and CosmoNest results appears.
It is worth to note that, the similar model with suppression on the large 
scales was shown as an example of use of the CosmoClust code 
\cite{Shaw:2007jj}. The bimodality of the likelihood 
functions was also observed and applicability of the CosmoClust code 
to that cases was emphasized.

The issue of constraining the bouncing cosmology with the observational data
was raised before in literature. In particular, studies based on SNIa data, location 
of acoustic peaks in the CMB and constraints from primordial nucleosynthesis (BBN)
were performed in Ref. \cite{Szydlowski:2005qb,Szydlowski:2008zz}. However, these 
cosmographic methods are inefficient in searching for the effects of the bounce. 
It is due to the fact that the factor $\frac{\rho}{\rho_{\text{c}}}$ is extremely low 
at the energy scales covered with this method.  Even during the BBN, where 
$T_{\text{BBN}} \sim 1$ MeV, we have $\rho_{\text{BBN}} \approx 10^{-90} 
\rho_{\text{Pl}}$. Therefore, if  $\rho_{\text{c}}\approx \rho_{\text{Pl}}$, we have 
$\frac{\rho}{\rho_{\text{c}}}\approx 10^{-90}$ and the holonomy corrections in the
Friedmann equation (\ref{Friedmann}) are vanishingly small \footnote{The constraint 
from  the BBN can be however more significant in case of the so-called inverse 
volume effects in LQC \cite{Bojowald:2007pc}}. Based on the method developed
in the present paper, we reach $\rho_{\text{obs}} = \frac{m^2\phi^2_{\text{obs}}}{2}
\approx 10^{-11}  \rho_{\text{Pl}}$, what gives $\frac{\rho}{\rho_{\text{c}}}\approx 10^{-11}$ for
$\rho_{\text{c}}\approx \rho_{\text{Pl}}$. Therefore, sensitivity on the holonomy corrections 
was increased around $10^{80}$ times with respect to the BBN constraint.  

Based the the results presented in this section one can conclude that the Big Bounce 
is consistent with the observations up to energy scales $\approx 10^{-11}  \rho_{\text{Pl}}$.
In this region the Big Bounce and Big Bang cosmologies are indistinguishable in the 
light of the available observational data. The advantage of the Big Bounce model is
however that the initial singularity problem is resolved and the initial conditions for the 
phase of inflation are naturally generated.

\section{Can we see the Big Bounce?} \label{BBounce}

The present value of scale $\lambda_*$ is crucial from the point of possible 
observational investigations of the Big Bounce cosmology. As it was discussed 
before, this scale overlaps with the size of the Hubble radius at the beginning 
of inflation. Therefore, it corresponds to the point of maximal displacement of the inflaton 
field, namely $\phi_{\text{max}}$. In this section we investigate how the variation 
of $\phi_{\text{max}}$ influences on the present value of $\lambda_*$. Based on 
this, it will be possible to investigate the observational conditions on the bounce.

In Fig. \ref{phimax}, the schematic illustration of the scalar field evolution near the place of 
the maximal displacement was shown.
In this figure we have marked the discussed 
$\phi_{\text{max}}$ value as well as the observed value $\phi_{\text{obs}}=2.9 m_{\text{Pl}}$.
While $\phi=\phi_{\text{obs}}$, the modes of the present size $\lambda_0 = 3.14$ Gpc (pivot scale)
were formed. Based on this, we can determinate what is the present size of the mode, which was 
equal to the Hubble radius at $\phi=\phi_{\text{max}}$. The transition from $\phi=\phi_{\text{max}}$
to $\phi \approx 0$ corresponding to the total amount of  $e$-foldings from inflation, which
can be decomposed as follows $N_{\text{tot}} = \Delta N + N_{\text{obs}}$. 
Here $N_{\text{obs}}$ is the observed value which corresponds 
to the transition from $\phi=\phi_{\text{obs}}$ to $\phi \approx 0$.
The number of $e$-foldings during the transition from  $\phi_{\text{max}}$ to $\phi_{\text{obs}}$
can be expressed as follows
\begin{eqnarray}
\Delta N = -\frac{4\pi}{m^2_{\text{Pl}}} \int_{\phi_{\text{max}}}^{\phi_{\text{obs}}} \frac{V}{V'} d\phi
=\frac{2\pi}{m^2_{\text{Pl}}} \left(\phi^2_{\text{max}}- \phi^2_{\text{obs}} \right). 
\end{eqnarray}

Based on this expression as well as on the Friedmann equation, the present value of $\lambda_*$ can be expressed as follows 
\begin{eqnarray}
\lambda_* = \lambda_0 \left( \frac{\phi_{\text{max}}}{m_{\text{Pl}}} \right)
\left( \frac{m_{\text{Pl}}}{\phi_{\text{obs}}} \right)  
\exp\left\{2\pi\left( \frac{\phi_{\text{max}}}{m_{\text{Pl}}} \right)^2-
 2\pi\left( \frac{\phi_{\text{obs}}}{m_{\text{Pl}}} \right)^2\right\}, \label{lstar}
\end{eqnarray}
where $\lambda_0 = 3.14$ Gpc and  $\phi_{\text{obs}}=2.9 m_{\text{Pl}}$. In Fig. \ref{lstarfig} we plot function
$\lambda_* (\phi_{\text{max}})$ given by (\ref{lstar}). For comparison, we also show some relevant length scales.
\FIGURE{
\centering
\includegraphics[width=8cm,angle=0]{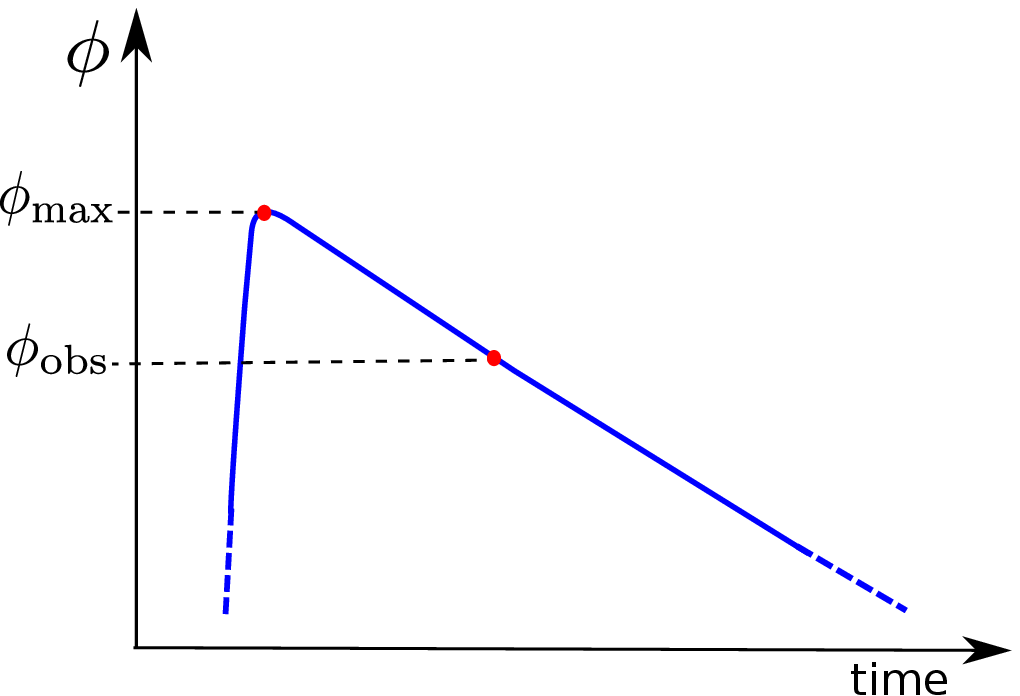}
\caption{Schematic illustration of the scalar field evolution near the place of 
the maximal displacement. The $\phi_{\text{max}}$ is a maximal displacement 
of the field. The $\phi_{\text{obs}}$ is the value of the scalar field that corresponds
to the powers spectrum measured at the pivot scale $\lambda_0 = 3.14$ Gpc.}
\label{phimax}
}
The first one is the Hubble radius $H_0/c \approx 4$ Gpc. The second is the distance to last scattering 
shell (LSS), $D_{\text{LSS}} \approx 14$ Gpc.  The last scale is the scale of suppression 
$\lambda_* \approx 40$ Gpc obtained in Sec.  \ref{CMB}. If $\phi_{\text{max}} > 2.94 m_{\text{Pl}}$
then the scale $\lambda_{*}$ is placed behind the scale of LSS. In such a case there is no 
chance to see the effect of suppression directly. It is because, the scale of suppression 
is higher than the physical horizon of photons, released during the recombination. Therefore 
only if $\phi_{\text{max}} < 2.94 m_{\text{Pl}}$, there is a possibility to study the effects of 
suppression on the CMB. From the fit performed in  Sec.  \ref{CMB} we got $\lambda_* \approx 40$ 
Gpc, what correspond to  $\phi_{\text{max}} \approx 2.97 m_{\text{Pl}}$.  Based on this, the one 
particular evolutionary trajectory can be distinguished. However, one have to keep in mind that 
the probability distribution on the parameter $k_*$ was unbounded from below. Therefore the 
obtained value $\phi_{\text{max}} \approx 2.97 m_{\text{Pl}}$ could be seen rather as a lower constraint
on $\phi_{\text{max}}$. As mentioned, in order to make the direct observations of the 
suppression possible, the value of  $\phi_{\text{max}}$ should be smaller than $2.94 m_{\text{Pl}}$. 
The observations suggest that this value is higher, what unfortunately exclude this possibility. 
Based on this one can however  exclude some models, where the predicted value of $\phi_{\text{max}}$
is not higher than $2.94 m_{\text{Pl}}$. This is in fact a case for the \emph{symmetric} inflation as studied in Ref. \cite{Chiou:2010nd}.
The issue of constraining this model was preliminary discussed in \cite{Mielczarek:2010bh}.
This is also still possible that the effect of oscillations can be observed. Perhaps it is even the 
reason why the particular value of  $k_*$ was distinguished from the WMAP observations. Namely, 
it was possible because the structure of modulations at the low multipoles was reconstructed, 
not because the scale of suppression was detected.
\FIGURE{
\centering
\includegraphics[width=8cm,angle=0]{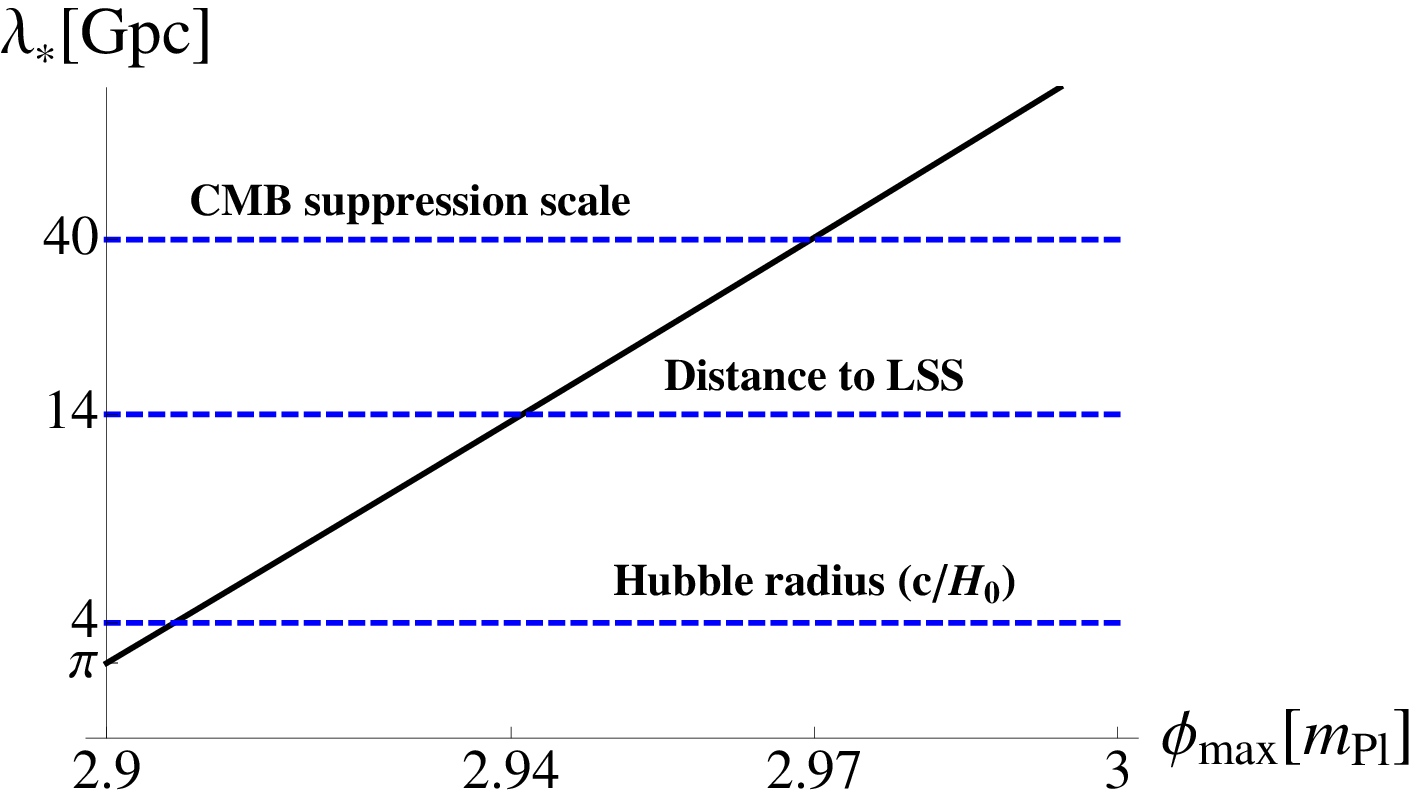}
\caption{The present value of the scale $\lambda_*$ as a function of $\phi_{\text{max}}$.}
\label{lstarfig}}

We finish this section with discussion of the observational constraint on
the the parameter $\rho_{\text{c}}$. In loop quantum cosmology, total energy 
density is constrained by $\rho \leq \rho_{\text{c}}$. At the stage of inflation 
where the present pivot scale structure were initiated, the energy density is equal to  
\begin{equation}
\rho_{\text{obs}} = \frac{m^2\phi^2_{\text{obs}} }{2} \approx 8 \cdot 10^{-12} m^4_{\text{Pl}}. \nonumber
\end{equation}
Based on this, we infer that $\rho_{\text{c}}>\rho_{\text{obs}}$.  Because
$\rho_{\text{obs}}  \ll  \rho_{\text{Pl}}$, the  observed constraint on the energy scale 
of the bounce is very weak. However, since $\rho_{\text{c}}\sim 1/\gamma^3$, the constraint 
on the parameter $\gamma$ can be much stronger. Indeed, based on (\ref{rhoc}) we find
\begin{equation}
\gamma < 1100. \label{gammaconstraint}
\end{equation}
The value obtained from consideration of black hole entropy $\gamma=0.239$ places
well within the observational bound. The constraint (\ref{gammaconstraint})  is quite
strong, however it must be kept in mind that it is based on relation (\ref{rhoc}), which 
can be invalid. As discussed in  Ref. \cite{Malkiewicz:2009zd}, the $\rho_{\text{c}}$ can be a free
parameter, and then it would be impossible to put the constraint as (\ref{gammaconstraint}).
Therefore, more theoretical predictions regarding the phenomenological parameters 
as, $\rho_{\text{c}}$,  are still awaiting.

To conclude, some models of the bouncing cosmology can be excluded  based on 
the observations of CMB. It is based on the observational constraint on $\phi_{\text{max}}$.
The direct observations of the bounce effects are however much harder to detect. 
As we have indicated, the effect of suppression cannot be used.  It is because the scale 
of suppression was shown to be higher than the scale of horizon. The effect of oscillations 
gives a chance, however the effect is, in general,  weaker and can be below the 
\emph{cosmic variance}. It must be also pointed out that the discussed effects can be also 
predicted from the different models. Therefore the important task is to find the observable 
which enables to distinguish between the models.

\section{Summary} \label{Summary} 

In this paper we have examined observations of the cosmic microwave 
background radiation as the potential probe of physics in the Planck epoch.
We have based our considerations on the predictions of loop quantum 
cosmology.  Within this approach the initial singularity is replaced by a
cosmic bounce.  During the bounce, the universe reaches the maximal energy
density $\rho_{\text{c}} < \infty$. In our studies, we have concentrated on the 
model with a massive scalar field. The advantage of this choice is that 
the phase of the standard slow-roll inflation is realized during the evolution.
During this phase, the primordial perturbations are formed. Their properties 
can be investigated by observations of the CMB. Based on the recent results
from the seven years of observations made by the WMAP satellite, we have
determined some parameters of the model. In particular, we have found that 
inflaton mass $m=(1.4 \pm 0.5) \cdot 10^{-6} m_{\text{Pl}}=(2.6 \pm 0.6) \cdot
10^{13}$ GeV. 

Subsequently, we have investigated the modifications  of the primordial power 
spectrum due to presence of the bounce.  The two main effects that were 
discussed are: suppression and oscillations of the inflationary spectrum. 
The suppression can very strongly modify the spectrum. However, the present
scale of suppression was shown to be behind the Hubble radius. Therefore, the effect
of suppression is not directly observationally available.  Despite this, the present scale of 
suppression $\lambda_*=\frac{2\pi}{k_*}  \approx 40 \ \text{Gpc}$ was 
distinguished by the observations. This is because of the oscillations in the
primordial power spectrum. The effect of oscillations is in general, more subtle 
and dominated by the cosmic variance. However, the present work indicates 
that the oscillations in the power spectrum can in fact explain the strong 
additional modulations in the spectrum of CMB on the low multipoles.  
In order to verify it, we have compared the Big Bounce model with the standard Big Bang 
scenario and showed  that the present observational data is not informative 
enough to distinguish these models. In other words, the Big Bounce predictions
were shown not to be  in conflict with the observational data.  Moreover, despite 
the Big Bounce model has one more parameter $k_*$, the evidence obtained in this case is 
comparable with the Big Bang case. 
 
We have shown that $\phi_{\text{max}}$ is not lower  than $2.97 m_{\text{Pl}}$. 
Based on this, some models of the bounce, as the mentioned \emph{symmetric}
model, can be  significantly constrained or even excluded. Assuming validity 
of (\ref{rhoc}) we gave an observational constraint on the Barbero-Immirzi parameter. 
We found that $\gamma < 1100$. The corresponding constraint on $\rho_{\text{c}}$
was shown to be much weaker. 

\acknowledgments

JM has been supported  by Polish Ministry of Science 
and Higher Education grant N N203 386437 and by 
Foundation of Polish Science award START.

\end{document}